\documentclass[12pt]{article}
\usepackage[T2A]{fontenc}
\usepackage[english]{babel}

\usepackage{latexsym}
\usepackage{amssymb}
\usepackage{amsmath}

\def\proof{\medbreak\noindent{\bf Proof}}

\newtheorem{theorem}{Theorem}
\newtheorem{lemma}{Lemma}

\def\{{\lbrace}
\def\}{\rbrace}

\def\cl{{\cal C}\!\ell}

\def\tr{{\rm tr}}

\def\C{{\Bbb C}}

\def\Tr{{\rm Tr}}

\def\det{{\rm det}}

\def\be{\begin{equation}}
\def\ee{\end{equation}}
\def\Det{{\rm Det}}

\def\Even{{\rm Even}}
\def\Odd{{\rm Odd}}
\newcommand{\mcc}[1]{\overleftarrow{#1}} % matrix complex conjugation

\newcommand{\st}{\stackrel}

\begin{document}

\title{Concepts of trace, determinant and inverse of Clifford algebra elements}

\author{D.~S.~Shirokov\\\\Steklov Mathematical Institute\\
Gubkin St.8, 119991 Moscow, Russia\\\\email: shirokov@mi.ras.ru}
\maketitle
\begin{abstract}
In our paper we consider the notion of determinant of Clifford algebra elements. We present some new formulas for determinant of Clifford algebra elements for the cases of dimension $4$ and $5$. Also we consider the notion of trace of Clifford algebra elements. We use the generalization of the Pauli's theorem for 2 sets of elements that satisfy the main anticommutation conditions of Clifford algebra.
\end{abstract}


MSC classes: 	15A66

%\tableofcontents

\newpage

\section{Introduction}

The notion of determinant of Clifford algebra elements was considered in \cite{MarcMart}. In our work we present some new formulas for determinant of Clifford algebra elements for the cases of dimension $n=4$ and $5$. Also we consider the notion of trace of Clifford algebra elements. We use the generalization of the Pauli's theorem for 2 sets of elements that satisfy the main anticommutation conditions of Clifford algebra.

After writing this paper author found the article \cite{new} on the subject that is close to the subject of this paper. In particular, the article \cite{new} contains the formulas that are similar to the formulas for the determinant in this paper. However, note that for the first time most of these formulas ($n=1, 2, 3$) were introduced in \cite{MarcMart}.

\bigskip

%%%%%%%%%%%%%%%%%%%%%%%%%%%%%%%%%%%%%%%%%%%%%%%%%%%%%%%%%%%%%%%%%%%%%%%%%%%%%%
%%%%%%%%%%%%%%%%%%%%%%%%%%%%%%%%%%%%%%%%%%%%%%%%%%%%%%%%%%%%%%%%%%%%%%%%%%%%%%

\section{Complex Clifford algebras}

Let $p$ and $q$ be nonnegative integers such that $p+q=n\geq 1$. We consider complex Clifford algebra $\cl(p,q)$.
The construction of Clifford algebra $\cl(p,q)$ is discussed in details in \cite{Marchuk:Shirokov}.

Generators
$e^1, e^2, \ldots, e^n$ satisfy the following conditions
$$e^a e^b +e^b e^a = 2\eta^{ab} e,$$
where $\eta=||\eta^{ab}||$ is the diagonal matrix whose diagonal contains $p$ elements equal to $+1$ and $q$ elements equal to $-1$.

The elements
$$e^{a_1}\ldots e^{a_k}=e^{a_1\ldots a_k},\qquad 1\leq a_1<\ldots a_k\leq n,\quad k=1, 2, \ldots n$$
together with the identity element $e$ form a basis of Clifford algebra $\cl(p,q)$. The number of basis elements equals to $2^n$.

Any Clifford algebra element $U\in\cl(p,q)$ can be written in the following form
\begin{eqnarray}
U=ue+u_a e^a+\sum_{a_1<a_2}u_{a_1 a_2}e^{a_1 a_2}+\ldots+u_{1\ldots n}e^{1\ldots n},\label{U:decomp}
\end{eqnarray}
where $u, u_a, u_{a_1 a_2},\ldots u_{1\ldots n}$ are complex constants.

We denote the vector subspaces spanned by the elements $e^{a_1 \ldots a_k}$ enumerated by the ordered multi-indices of length $k$ by $\cl_k(p,q)$. The elements of the subspace $\cl_k(p,q)$ are denoted by $\st{k}{U}$ and called elements of rank $k$. We have
\begin{eqnarray}
\cl(p,q)=\oplus_{k=0}^{n}\cl_k(p,q).\label{ranks}
\end{eqnarray}

Clifford algebra $\cl(p,q)$ is a superalgebra, so we have even and odd subspaces:
\begin{eqnarray}
\cl(p,q)=\cl_{\Even}(p,q)\oplus\cl_{\Odd}(p,q),\label{Evenness}
\end{eqnarray}
where
$$\cl_{\Even}(p,q)=\cl_0(p,q)\oplus\cl_2(p,q)\oplus\cl_4(p,q)\oplus\ldots,$$
$$\cl_{\Odd}(p,q)=\cl_1(p,q)\oplus\cl_3(p,q)\oplus\cl_5(p,q)\oplus\ldots$$

\section{Operations of conjugation}

Let denote complex conjugation of matrix by $\mcc{A}$, transpose matrix by $A^T$, Hermitian conjugate matrix (composition of these 2 operations)
by $A^\dagger$.

Now let define some operations on Clifford algebra elements.

\noindent{\bf Complex conjugation.} Operation of complex conjugation $U\to\bar{U}$ acts in the following way
\begin{eqnarray}
\bar U=\mcc{u}e+\mcc{u_a} e^a+\sum_{a_1<a_2}\mcc{u_{a_1 a_2}}e^{a_1
a_2}+\sum_{a_1<a_2<a_3}\mcc{u_{a_1 a_2 a_3}}e^{a_1 a_2
a_3}+\ldots\label{U:decomp1}
\end{eqnarray}
We have
$$\bar e^a=e^a,\quad a=1, \ldots, n,\qquad \overline{\overline U}=U,\qquad (\overline{U V})=\bar U \bar V,\qquad (\overline{U+V})=\bar U+\bar V,$$
$$(\overline{\lambda U})=\mcc{\lambda} \bar U,\qquad \forall U,V\in \cl(p,q),\qquad \lambda\in\C.$$

\noindent{\bf Reverse.} Let define operation reverse for $U\in\cl(p,q)$ in the following way
\begin{eqnarray*}
U^\sim &=& \sum_{k=0}^n(-1)^{\frac{k(k-1)}{2}}\st{k}{U}.
\end{eqnarray*}
We have
$$(e^a)^\sim=e^a,\quad a=1,\ldots,n,\qquad U^{\sim\sim}=U,\qquad (U V)^\sim=V^\sim U^\sim,$$
$$(U+V)^\sim=U^\sim+V^\sim,\qquad (\lambda U)^\sim=\lambda U^\sim.$$

\noindent{\bf Pseudo-Hermitian conjugation.} Let define Pseudo-Hermitian conjugation as composition of reverse and complex conjugation:
$$
U^\ddagger=\bar U^\sim.
$$
We have
$$(e^a)^\ddagger=e^a,\qquad a=1,\ldots,n,\qquad U^{\ddagger\ddagger}=U,\qquad (U V)^\ddagger=V^\ddagger U^\ddagger,$$
$$(U+V)^\ddagger=U^\ddagger+V^\ddagger, \qquad (\lambda U)^\ddagger=\mcc{\lambda} U^\ddagger.$$

\noindent{\bf Grade involution.} Let define operation of grade involution $U\to U^\curlywedge$ in the following way
$$
U^\curlywedge=\sum_{k=0}^n(-1)^k\st{k}{U}.
$$
We have
$$(e^a)^\curlywedge=-e^a,\quad a=1,\ldots,n,\qquad U^{\curlywedge\curlywedge}=U,\qquad (U V)^\curlywedge=U^\curlywedge V^\curlywedge,$$
$$(U+V)^\curlywedge=U^\curlywedge+V^\curlywedge,\qquad (\lambda U)^\curlywedge=\lambda U^\curlywedge.$$

\noindent{\bf Clifford conjugation.}
Let define Clifford conjugation as composition of grade involution and reverse $U\to
U^{\curlywedge\sim}$:
$$
U^{\curlywedge\sim}=\sum_{k=0}^n(-1)^{\frac{k(k+1)}{2}}\st{k}{U}.
$$
We have
$$(e^a)^{\curlywedge\sim}=-e^a,\quad a=1,\ldots,n,\qquad U^{\curlywedge\sim\curlywedge\sim}=U,\qquad (U V)^{\curlywedge\sim}=V^{\curlywedge\sim}U^{\curlywedge\sim},$$
$$(U+V)^{\curlywedge\sim}=U^{\curlywedge\sim}+V^{\curlywedge\sim},\qquad (\lambda U)^{\curlywedge\sim}=\lambda U^{\curlywedge\sim}.$$

\noindent{\bf Hermitian conjugation.}
In \cite{Marchuk:Shirokov} we consider operation of Hermitian conjugation. We have the following formulas for these operation:
\begin{eqnarray}
U^\dagger&=&(e^{1\ldots p})^{-1}U^\ddagger e^{1 \ldots p},\qquad \mbox{if $p$ - odd},\nonumber\\
U^\dagger&=&(e^{1\ldots p})^{-1}U^{\ddagger\curlywedge} e^{1 \ldots p},\qquad \mbox{if $p$ - even},\\
U^\dagger&=&(e^{p+1\ldots n})^{-1}U^\ddagger e^{p+1 \ldots n},\qquad \mbox{if $q$ - even},\nonumber\\
U^\dagger&=&(e^{p+1\ldots n})^{-1}U^{\ddagger\curlywedge} e^{p+1 \ldots n},\qquad \mbox{if $q$ - odd},\nonumber
\end{eqnarray}
We have
$$(e^a)^\dagger=(e^a)^{-1},\qquad a=1,\ldots,n,\qquad U^{\dagger\dagger}=U,\qquad (U V)^\dagger=V^\dagger U^\dagger,$$
$$(U+V)^\ddagger=U^\dagger+V^\dagger, \qquad (\lambda U)^\dagger=\mcc{\lambda} U^\dagger.$$

\section{Matrix representations of Clifford algebra elements, recurrent method.}

Complex Clifford algebras $\cl(p,q)$ of dimension $n$ and different signatures $(p,q), p+q=n$ are isomorphic. Clifford algebras $\cl(p,q)$ are isomorphic to the matrix algebras of complex matrices. In the case of even $n$ these matrices are of order $2^{\frac{n}{2}}$. In the case of odd $n$ these matrices are block diagonal of order $2^{\frac{n+1}{2}}$ with 2 blocks of order $2^{\frac{n-1}{2}}$.

Consider the following matrix representations of Clifford algebra elements.

Identity element $e$ of Clifford algebra $\cl(p,q)$ maps to identity matrix of corresponding order:
$e \to {\bf 1}$.

For $\cl(1,0)$ element $e^1$ maps to the following matrix
$$
e^1\to\left( \begin{array}{ll}
 1 & 0 \\
 0 & -1 \end{array}\right).
$$
For $\cl(2,0)$ we have
$$
e^1\to\left( \begin{array}{ll}
 1 & 0 \\
 0 & -1 \end{array}\right)
,\quad
e^2\to\left( \begin{array}{ll}
 0 & 1 \\
 1 & 0 \end{array}\right).
$$
Further, suppose we have a matrix representation for $\cl(2k,0)$, $n=2k$:
$$e^1, \ldots, e^n \to \gamma^1, \ldots, \gamma^n.$$

Then, for Clifford algebra $\cl(2k+1,0)$ we have
$$
e^a\to\left( \begin{array}{ll}
 \gamma^a & 0 \\
 0 & -\gamma^a \end{array}\right)
,\quad a=1, \ldots, n,\qquad
e^{n+1}\to\left( \begin{array}{ll}
 i^k \gamma^1\ldots \gamma^n & 0 \\
 0 & -i^k \gamma^1\ldots \gamma^n \end{array}\right).
$$

For Clifford algebra $\cl(2k+2,0)$ we have the same matrices for $e^a, a=1, \ldots, n+1$ as in the previous case $n=2k+1$ and for $e^{n+2}$ we have
$$
e^{n+2}\to\left( \begin{array}{ll}
 0 & {\bf1} \\
 {\bf1} & 0 \end{array}\right).
$$

So, we have matrix representation for all Clifford algebras $\cl(n,0)$. In the cases of other signatures elements $e^a, a>p$ maps to the same matrices as in signature $(n,0)$ but with multiplication by imaginary unit $i$.

For example, we have the following matrix representations for Clifford algebras $\cl(3,0)$, $\cl(4,0)$ and $\cl(1,3)$.

\begin{description}
  \item[$\cl(3,0)$:]
$$
  e^1 \to \gamma^1=\left( \begin{array}{llll}
 1 & 0 & 0 & 0\\
 0 & -1 & 0 & 0\\
 0 & 0 & -1 & 0\\
 0 & 0 & 0 & 1\end{array}\right),\quad
  e^2 \to \gamma^2=\left( \begin{array}{llll}
 0 & 1 & 0 & 0\\
 1 & 0 & 0 & 0\\
 0 & 0 & 0 & -1\\
 0 & 0 & -1 & 0\end{array}\right),$$
 $$e^3 \to \gamma^3=\left( \begin{array}{llll}
 0 & i & 0 & 0\\
 -i & 0 & 0 & 0\\
 0 & 0 & 0 & -i\\
 0 & 0 & i & 0\end{array}\right).
 $$
  \item[$\cl(4,0)$:]
  $$e^1 \to \gamma^1,\quad e^2 \to \gamma^2,\quad e^3 \to \gamma^3,\quad e^4 \to \gamma^4=\left( \begin{array}{llll}
 0 & 0 & 1 & 0\\
 0 & 0 & 0 & 1\\
 1 & 0 & 0 & 0\\
 0 & 1 & 0 & 0\end{array}\right).
 $$
 \item[$\cl(1,3)$:]
  $$e^1 \to \gamma^1,\quad e^2 \to i\gamma^2,\quad e^3 \to i\gamma^3,\quad e^4 \to i\gamma^4.
  $$
\end{description}

\section{Operation of trace of Clifford algebra elements}

Consider complex Clifford algebra $\cl(p,q)$ and introduce the operation of {\em trace} of Clifford algebra element $U\in\cl(p,q)$ as the following operation of projection onto subspace $\cl_0(p,q)$:
\begin{equation}
\Tr(U)=\langle U\rangle_0|_{e\to 1}.\label{tracedef}
\end{equation}
For arbitrary element $U\in\cl(p,q)$ in the form (\ref{U:decomp}) we have
$$\Tr(ue+u_a e^a +\ldots)=u.$$

\begin{theorem}. \label{theoremTraceProp} Operation trace (\ref{tracedef}) of Clifford algebra element $U\in\cl(p,q)$ has the following properties:
\begin{itemize}
  \item linearity: $$\Tr(U+V)=\Tr(U)+\Tr(V),\qquad \Tr(\alpha U)=\alpha \Tr(U)$$ $$\forall U, V\in\cl(p,q),\qquad \forall \alpha\in\C,$$
  \item cyclic recurrence: $$\Tr(UV)=\Tr(VU),\qquad \Tr(UVW)=\Tr(VWU)=\Tr(WUV)$$ $$\forall U, V, W\in\cl(p,q),$$
  but, in general: $$\Tr(UVW)\neq\Tr(UWV).$$
  \item invariance under similarity transformation: $$\Tr(U^{-1}VU)=\Tr(V) \qquad \forall V\in\cl(p,q),\, U\in\cl^{\times}(p,q),$$
  where $\cl^{\times}(p,q)$ is the set of all invertible Clifford algebra elements.
  \item invariance under conjugations: $$\Tr(U)=\Tr(U^\curlywedge)=\Tr(U^\sim)=\mcc{\Tr(\overline{U})}=\mcc{\Tr(U^\ddagger)}=\mcc{\Tr(U^\dagger)}.$$
\end{itemize}

\end{theorem}

\proof. \, Linearity follows from the definition (\ref{tracedef}).

We have $$\Tr(UV)=\Tr(VU)$$ because $\Tr([\st{k}{U},\st{l}{V}])=0$ for $k,l=0,\ldots n$ (see \cite{Marchuk:Shirokov}). Then we obtain cyclic recurrence for 3 elements. We obtain invariance under similarity transformation as a simple consequence of cyclic recurrence. Last properties follow from properties of conjugations. $\blacksquare$

There is a relation between operation trace $\Tr$ of Clifford algebra element $U\in \cl(p,q)$ and operation trace $\tr$ of quadratic matrix.
To obtain this relation, at first, we will prove the following statement.

\begin{lemma}. \label{lemmaTrace}
Consider recurrent matrix representation of Clifford algebra $\cl(p,q)$ (see above). For this representation $U \to \underline U$ we have
$$\tr(\underline U)=2^{[\frac{n+1}{2}]} \Tr(U),\qquad \tr(\underline U^\curlywedge)=\tr(\underline U).$$

\end{lemma}

\proof. \, Coefficient $2^{[\frac{n+1}{2}]}$ equals to the order of corresponding matrices. It is not difficult to see that trace of almost all matrices that correspond to basis elements equals to zero $$\tr(\underline e^A)=0,\qquad \mbox{where $A$ - any multi-index except empty}.$$
The only exception is identity element $e$, which corresponds to the identity matrix. In this case we have $\tr(\underline e)=2^{[\frac{n+1}{2}]}$. Further we use linearity of trace and obtain $$\tr(\underline U)=2^{[\frac{n+1}{2}]} u=2^{[\frac{n+1}{2}]}\Tr(U).$$
The second property is a simple consequence of the first property, because
$$\tr(\underline U^\curlywedge)=2^{[\frac{n+1}{2}]} \Tr(U^\curlywedge)=2^{[\frac{n+1}{2}]} \Tr(U)=\tr(\underline U).$$
$\blacksquare$

\begin{theorem}. \label{theoremTraceDef}
Consider complex Clifford algebra $\cl(p,q)$ and operation trace $\Tr$. Then
\begin{eqnarray}
\Tr(U)=\frac{1}{2^{[\frac{n+1}{2}]}}\tr(\gamma(U)),\label{tracedef2}
\end{eqnarray}
where $\gamma(U)$ - any matrix representation of Clifford algebra $\cl(p,q)$ of minimal dimension. Moreover, this definition of trace (\ref{tracedef2}) is equivalent to the definition (\ref{tracedef}). New definition is well-defined because it doesn't depend on the choice of matrix representation.

\end{theorem}

\proof. \, This property proved in the previous statement for the recurrent matrix representation. Let we have besides recurrent matrix representation $$\underline U=U|_{e^a\to\gamma^a}$$ another matrix representation $$\underline{\underline U}=U|_{e^a\to\beta^a}.$$ Then, by Pauli's theorem in Clifford algebra of even dimension $n$ there exists matrix $T$ such that $$\beta^a=T^{-1}\gamma^a T,\qquad a=1, \ldots, n.$$ Then, we have
$$\underline{\underline U}= T^{-1} \underline UT$$ and
$$\tr(\underline{\underline U})= \tr(T^{-1}\underline U T)=\tr(\underline U).$$
In the case of odd $n$ we can have also another case (by Pauli's theorem), when two sets of matrices relate in the following way
$$\beta^a=-T^{-1}\gamma^a T,\qquad a=1, \ldots, n.$$ In this case we have
$$\underline{\underline{U}}= T^{-1}\underline{U^\curlywedge}T.$$
From $\tr(\underline{U^\curlywedge})=\tr(\underline U)$ (see Lemma \ref{lemmaTrace}) we obtain
$$\tr(\underline{\underline{U}})= \tr(T^{-1}\underline{U^\curlywedge}T)=\tr(\underline{U^\curlywedge})=\tr(\underline U).$$ $\blacksquare$

\section{Determinant of Clifford algebra elements}

{\it Determinant of Clifford algebra element $U\in\cl(p,q)$} is a complex number
\begin{eqnarray}\Det U=\det(\underline U),\label{det}\end{eqnarray}
which is a determinant of any matrix representation $\underline U$ of minimal dimension.

Now we want to show that this definition is well-defined. Let prove the following Lemma.

\begin{lemma}. \label{lemmaDetRek}
Consider the recurrent matrix representation (see above) of Clifford algebra $\cl(p,q)$. For this representation $U \to \underline U$ we have
$$\det(\underline U^\curlywedge)=\det(\underline U).$$

\end{lemma}

\proof. \, In the case of Clifford algebra of even dimension $n$ we have
$$U^\curlywedge=(e^{1\ldots n})^{-1}Ue^{1\ldots n}.$$
So, we obtain
$$\det(\underline{U^\curlywedge})=\det(\underline{(e^{1\ldots n})^{-1}U e^{1\ldots n}})=\det(\underline{(e^{1\ldots n})^{-1}})\det(\underline{U})\det(\underline{e^{1\ldots n}})=\det(\underline{U}).$$

In the case of Clifford algebra of odd dimension generators maps to the block diagonal matrices and blocks are identical up to the sign:
$$
e^a\to\left( \begin{array}{ll}
 \gamma^a & 0 \\
 0 & -\gamma^a \end{array}\right).
$$
Then for elements of the rank 2 we obtain
$$
e^{ab}\to\left( \begin{array}{ll}
 \gamma^a\gamma^b & 0 \\
 0 & \gamma^a\gamma^b \end{array}\right).
$$
It is not difficult to see that even part $U_{\Even}$ of arbitrary element $U=U_{\Even}+U_{\Odd}$ maps to the matrix with identical blocks, and odd part $U_{\Odd}$ of the element $U$ maps to the matrix with the blocks differing in sign:
$$
U_{\Even}\to\left( \begin{array}{ll}
 A & 0 \\
 0 & A \end{array}\right),\qquad
U_{\Odd}\to\left( \begin{array}{ll}
 B & 0 \\
 0 &-B \end{array}\right).
$$
Then we have
$$
U\to\left( \begin{array}{ll}
 A+B & 0 \\
 0 & A-B \end{array}\right),\qquad
U^\curlywedge\to\left( \begin{array}{ll}
 A-B & 0 \\
 0 & A+B \end{array}\right)
$$
and
$$\det(U)=(A-B)(A+B)=\det(U^\curlywedge).$$
$\blacksquare$

\begin{theorem}. \label{predlDet} Definition (\ref{det}) is well-defined, i.e. it doesn't depend on the matrix representation.

\end{theorem}

\proof. \, Consider the recurrent matrix representation $$\underline U=U|_{e^a\to\gamma^a}.$$
The statement for this representation proved in the previous lemma. Let we have another matrix representation $$\underline{\underline U}=U|_{e^a\to\beta^a}.$$ Then, by Pauli's theorem in Clifford algebra of even dimension $n$ there exists a matrix $T$ such that $$\beta^a=T^{-1}\gamma^a T,\qquad a=1, \ldots, n.$$ Then we have
$$\underline{\underline U}= T^{-1} \underline UT$$ and obtain
$$\det(\underline{\underline U})= \det(T^{-1}\underline U T)=\det(T^{-1})\det(\underline U)\det(T)=\det(\underline U).$$

In the case of odd $n$, by Pauli's theorem we also have another case, where 2 sets of matrices relate in the following way
$$\beta^a=-T^{-1}\gamma^a T,\qquad a=1, \ldots, n.$$ In this case we have
$$\underline{\underline{U}}= T^{-1}\underline{U^\curlywedge}T.$$
From $\det(\underline{U^\curlywedge})=\det(\underline U)$ (see Lemma \ref{lemmaDetRek}) we obtain
$$\det(\underline{\underline{U}})= \det(T^{-1}\underline{U^\curlywedge}T)=\det(\underline{U^\curlywedge})=\det(\underline U).$$ $\blacksquare$

Let formulate some properties of operation determinant of Clifford algebra element.

\begin{theorem}. \label{theoremDetProp} Operation determinant (\ref{det}) of Clifford algebra element $U\in\cl(p,q)$ has the following properties
\begin{itemize}
  \item \begin{equation}\Det(UV)=\Det(U)\Det(V),\qquad \Det(\alpha U)=\alpha^{2^{[\frac{n+1}{2}]}} \Det(U)\label{2pr}\end{equation} $$\forall U, V\in\cl(p,q),\qquad \forall \alpha\in\C.$$
  \item Arbitrary element $U\in \cl(p,q)$ is invertible if and only if $\Det U\neq 0$.
  \item For any invertible element $U\in\cl(p,q)$ \begin{equation}\Det (U^{-1}) = (\Det U)^{-1}.\label{3pr}\end{equation}
  \item Invariance under similarity transformation: $$\Det(U^{-1}VU)=\Det(V) \qquad \forall V\in\cl(p,q),\, U\in\cl^{\times}(p,q),$$
  where $\cl^{\times}(p,q)$ is set of all invertible Clifford algebra elements.
  \item Invariance under conjugations: $$\Det(U)=\Det(U^\curlywedge)=\Det(U^\sim)=\mcc{\Det(\overline{U})}=\mcc{\Det(U^\ddagger)}=\mcc{\Det(U^\dagger)}.$$
\end{itemize}

\end{theorem}

\proof. \, The first 4 properties are simple and follow from the definition of determinant (\ref{det}).

In lemma \ref{lemmaDetRek} we have the property $\Det(U)=\Det(U^\curlywedge)$ for the recurrent matrix representation. But it is also valid for other matrix representations because of independence on the choice of representations (see Theorem \ref{predlDet}).

It is known that operation $\sim$ relates to the operation of matrix transpose (as similarity transformation) and also we have $\det(U)=\det(U^T)$.
Analogously we can consider another operations of conjugation.
$\blacksquare$

Definition (\ref{det}) of determinant of Clifford algebra element $U\in\cl(p,q)$ is connected with its matrix representation. We have shown that this definition doesn't depend on matrix representation. So, determinant is a function of complex coefficients $u_{a_1 \ldots a_k}$ located before basis elements $e^{a_1\ldots a_k}$ in (\ref{U:decomp}). In the cases of small dimensions $n\leq 5$ we give expressions for determinant of Clifford algebra elements that doesn't relate to the matrix representation.

Now we need also 2 another operations of conjugations $\bigtriangledown$, $\bigtriangleup$:

$$(\st{0}{U}+\st{1}{U}+\st{2}{U}+\st{3}{U}+\st{4}{U})^\bigtriangledown = \st{0}{U}+\st{1}{U}+\st{2}{U}+\st{3}{U}-\st{4}{U},\qquad n=4,$$
$$(\st{0}{U}+\st{1}{U}+\st{2}{U}+\st{3}{U}+\st{4}{U}+\st{5}{U})^\bigtriangledown = \st{0}{U}+\st{1}{U}+\st{2}{U}+\st{3}{U}-\st{4}{U}-\st{5}{U},\qquad n=5,$$
$$(\st{0}{U}+\st{1}{U}+\st{2}{U}+\st{3}{U}+\st{4}{U}+\st{5}{U})^\bigtriangleup = \st{0}{U}+\st{1}{U}+\st{2}{U}+\st{3}{U}+\st{4}{U}-\st{5}{U},\qquad n=5.$$

\begin{theorem}. \label{theoremDet} We have the following formulas for the determinant of Clifford algebra element $U\in\cl(p,q)$:
\begin{eqnarray}
\Det\,U=\left\lbrace
\begin{array}{ll}
U, & n=0;\\
U U^\curlywedge, & n=1;\\
U U^{\sim\curlywedge}, & n=2;\\
U U^\sim U^\curlywedge U^{\sim\curlywedge}= U U^{\sim\curlywedge} U^\curlywedge U^\sim, & n=3;\\
U U^\sim (U^\curlywedge U^{\sim\curlywedge})^\bigtriangledown= U U^{\sim\curlywedge} (U^\curlywedge U^{\sim})^\bigtriangledown, & n=4;\\
U U^\sim (U^\curlywedge U^{\sim\curlywedge})^\bigtriangledown (U U^\sim (U^\curlywedge U^{\sim\curlywedge})^\bigtriangledown)^\bigtriangleup, & n=5.
\end{array}
\right.
\end{eqnarray}

\end{theorem}

Note, that these expressions are Clifford algebra elements of the rank $0$. In this case we identify them with the constants: $ue\equiv u$.

\proof.\  The proof is by direct calculation. $\blacksquare$.

Note, that properties (\ref{2pr}) and (\ref{3pr}) for small dimensions also can be proved with the formulas from Theorem \ref{theoremDet}. For example, in the case $n=3$ we have
$$\Det(UV)=UV (UV)^{\sim\curlywedge} (UV)^\curlywedge (UV)^\sim  =UVV^{\sim\curlywedge} U^{\sim\curlywedge}U^\curlywedge V^\curlywedge V^\sim U^\sim  =$$
$$=UU^{\sim\curlywedge} U^\curlywedge U^\sim   V V^{\sim\curlywedge} V^\curlywedge V^\sim =\Det(U)\Det(V)$$
We used the fact that $VV^{\sim\curlywedge}$ and $V^\curlywedge V^\sim=(VV^{\sim\curlywedge})^\sim$ are in Clifford algebra center $\cl_0(p,q)\oplus\cl_3(p,q)$ and commute with all elements.

Theorem \ref{theoremDet} give us explicit formulas for inverse in $\cl(p,q)$. We have the following theorem.

\begin{theorem}. \label{theoremObrat} Let $U$ be invertible element of Clifford algebra $\cl(p,q)$. Then we have the following expressions for $U^{-1}$:
\begin{equation}
(U)^{-1}=\left\lbrace
\begin{array}{ll}
\frac{e}{U}, & n=0;\\ \\
\frac{U^\curlywedge}{U U^\curlywedge}, & n=1;\\ \\
\frac{U^{\sim\curlywedge}}{U U^{\sim\curlywedge}}, & n=2;\\ \\
\frac{U^\sim U^\curlywedge U^{\sim\curlywedge}}{U U^\sim U^\curlywedge U^{\sim\curlywedge}}=\frac{U^{\sim\curlywedge} U^\curlywedge U^\sim}{U U^{\sim\curlywedge} U^\curlywedge U^\sim}, & n=3;\\ \\
\frac{U^\sim (U^\curlywedge U^{\sim\curlywedge})^\bigtriangledown}{U U^\sim (U^\curlywedge U^{\sim\curlywedge})^\bigtriangledown}=\frac{U^{\sim\curlywedge} (U^\curlywedge U^{\sim})^\bigtriangledown}{ U U^{\sim\curlywedge} (U^\curlywedge U^{\sim})^\bigtriangledown}, & n=4;\\ \\
\frac{U^\sim (U^\curlywedge U^{\sim\curlywedge})^\bigtriangledown (U U^\sim (U^\curlywedge U^{\sim\curlywedge})^\bigtriangledown)^\bigtriangleup}{U U^\sim (U^\curlywedge U^{\sim\curlywedge})^\bigtriangledown (U U^\sim (U^\curlywedge U^{\sim\curlywedge})^\bigtriangledown)^\bigtriangleup}, & n=5.
\end{array}
\right.
\end{equation}

\end{theorem}

Note, that in denominators we have Clifford algebra elements of the rank $0$. We identify them with the constants: $ue\equiv u$.

\proof.\ Statement follows from Theorem \ref{theoremDet}. $\blacksquare$

Note, that formulas for determinant in Theorem \ref{theoremDet} are not unique. For example, in the case of $n=4$ we can use the following formulas, but only for even and odd Clifford algebra elements $U\in\cl_{\Even}(p,q)\cup\cl_{\Odd}(p,q)$.

Consider operation $+$, that acts on the even elements such that it changes the sign before the basis elements that anticommutes with $e^1$. For example, elements $e, e^{23}, e^{24}, e^{34}$ maps under $+$ into themselves, and elements $e^{12}, e^{13}, e^{14}, e^{1234}$ change the sign.

\begin{theorem}. \label{theoremDet2} Let $U\in\cl_{\Even}(p,q)$, $n=p+q=4$. Then
\begin{equation}
\Det U= U U^\sim U^{\sim +} U^+.
\end{equation}
Let $U\in\cl_{\Odd}(p,q)$, $n=p+q=4$. Then
\begin{equation}
\Det U= U U^\sim U^{\sim} U.
\end{equation}

\end{theorem}

\proof.\, The proof is by direct calculation. $\blacksquare$

\end{document}